\documentclass{mem}
\usepackage{natbib}\usepackage{txfonts}\usepackage{balance}
\usepackage{graphicx}
\usepackage[a4paper]{hyperref}
\idline{75}{282}
\begin{document}
\def\teff{$T\rm_{eff }$}
\def\kms{$\mathrm {km s}^{-1}$}

\title{
The Curious Case of Abell 2256
}

   \subtitle{}

\author{
Tracy E.\, Clarke\inst{1}, 
Torsten En{\ss}lin\inst{2},
Alexis Finoguenov\inst{3,4},
Huib Intema\inst{5},
Christoph Pfrommer\inst{6},
Reinout van Weeren\inst{7},
Huub R\"ottgering\inst{7},
\and Raymond Oonk\inst{7}
          }

  \offprints{T. Clarke}

\institute{
Naval Research Laboratory
Remote Sensing Division, Code 7213
4555 Overlook Ave SW,
Washington, DC 20375 USA
\and
Max-Planck-Institut f\"ur Astrophysik, 
Karl-Schwarzschild-Str. 1, 
85741 Garching, Germany
\and
Max-Planck-Institut f\"ur Extraterrestrische Physik, 
Giessenbachstr., 
85748 Garching, Germany 
\and
University of Maryland, Baltimore County, 
1000 Hilltop Cr., 
Baltimore, MD 21250, USA
\and
National Radio Astronomy Observatory, 
Charlottesville, VA, USA
\and
Canadian Institute for Theoretical Astrophysics, 
University of Toronto, 60 St George Street, 
Toronto, Ontario M5S 3H8, Canada
\and
Leiden Observatory, 
Leiden University, 
P.O. Box 9513, 
NL-2300 RA Leiden, Netherlands
\email{tracy.clarke.ca@nrl.navy.mil}
}

\authorrunning{Clarke }

\titlerunning{Abell 2256}

\abstract{Abell 2256 is a rich, nearby (z=0.0594) galaxy cluster that
  has significant evidence of merger activity. We present new radio
  and X-ray observations of this system. The low-frequency radio
  images trace the diffuse synchrotron emission of the Mpc-scale radio
  halo and relics as well as a number of recently discovered, more
  compact, steep spectrum sources. The spectral index across the
  relics steepens from the north-west toward the south-east. Analysis
  of the spectral index gradients between low and and high-frequencies
  shows spectral differences away from the north-west relic edge such
  that the low-frequency index is significantly flatter than the high
  frequency spectral index near the cluster core. This trend would be
  consistent with an outgoing merger shock as the origin of the relic
  emission. New X-ray data from XMM-Newton reveal interesting
  structures in the intracluster medium pressure, entropy and
  temperature maps. The pressure maps show an overall low pressure
  core co-incident with the radio halo emission, while the temperature
  maps reveal multiple regions of cool emission within the central
  regions of Abell 2256. The two cold fronts in Abell 2256 both appear
  to have motion in similar directions.  } \maketitle{}

\section{Introduction}

Clusters of galaxies fill an exceptional role in astrophysics as both
laboratories for plasma physical processes as well as key constituents
in precision cosmology studies of dark energy. Major cluster mergers
are the most energetic events since the Big Bang and can drive up to
$10^{64}$ ergs of gravitational potential energy into the intracluster
medium \citep{sarazin02}. This energy is then released through shocks
and turbulence which heat the intracluster medium (ICM), accelerate
relativistic particles and compress magnetic fields. Merging clusters
are often (but not always) characterized by the presence of diffuse,
low-frequency synchrotron emission believed to be a direct result of
the particle acceleration and field compression. The radio sources are
characterized as {\it radio halos} and {\it relics} depending on the
observational characteristics \citep[see e.g.][]{kempner, ferrari}.

Combining X-ray imaging and spectroscopic data with detailed
multi-frequency radio total intensity and polarization studies for
individual merging systems can provide insight into particle
acceleration process as well as the number of recent mergers, their
age, and orientation. Here, we briefly discuss recent radio and X-ray
observations of the merging cluster Abell 2256. Previous radio studies
of Abell 2256 have revealed a Mpc scale halo in addition to the well
known relics \citep{clarke06, brentjens08, kale10}, while X-ray
studies \citep[see e.g.][]{sun} have revealed significant evidence of
a disturbed ICM.

\section{Steep Spectrum Radio Sources}

\begin{figure}
\resizebox{\hsize}{!}{\includegraphics[clip=true]{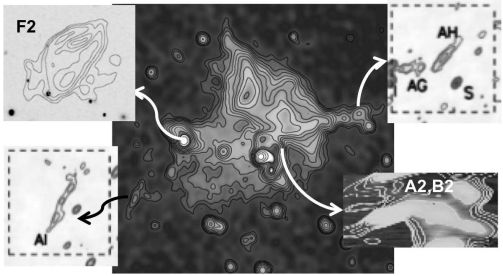}}
\caption{\footnotesize The central image shows an NRAO VLA 325 MHz
  radio image of Abell 2256. The insets show blow-ups of regions of
  steep spectrum radio emission. Clockwise from the top left: region
  F2 \citep{miller03}, regions AG \& AH \citep{vanweeren}, regions A2 \& B2
  \citep{intema}, region AI \citep{vanweeren}.}
\label{A2256_ss}
\end{figure}
Abell 2256 has recently been the target of many new deep low-frequency
high-resolution radio interferometric studies. These observations have
revealed a previously unknown population of $\sim$ 100 -- 200 kpc long
steep spectrum synchrotron filaments surrounding the cluster
center. In Figure~\ref{A2256_ss} we show a VLA 325 MHz radio image of
Abell 2256 (Clarke et al., in prep.) together with panels showing the
well-known ultra-steep spectrum source F2, and the newly revealed
filaments AG, AH, AI, A2 and B2. Sources AG, AH, and AI are located
nearly a Mpc in projection from the cluster center \citep{vanweeren}
and are classified by \citeauthor{vanweeren} as likely radio phoenix
sources \citep[see terminology in][]{kempner}.

The very steep spectral indices of the new synchrotron filaments AG,
AH, AI, A2 and B2 ($\alpha < -1.5$) suggest that they are all likely
due to merger shock induced adiabatic compression of fossil radio
plasma. The presence of this new population of synchrotron filaments
in Abell 2256 is not likely unique to this cluster but rather would be
expected to be a common property of merging clusters. These new
sources therefore provide a powerful new tool with which to trace
cluster merger shocks in deep maps from the next generation of
low-frequency interferometers such as LOFAR \citep{rottgering10} and
the Long Wavelength Array \citep{ellingson}.

\section{Radio Relic Spectral Index}

\begin{figure}
\resizebox{\hsize}{!}{\includegraphics[clip=true]{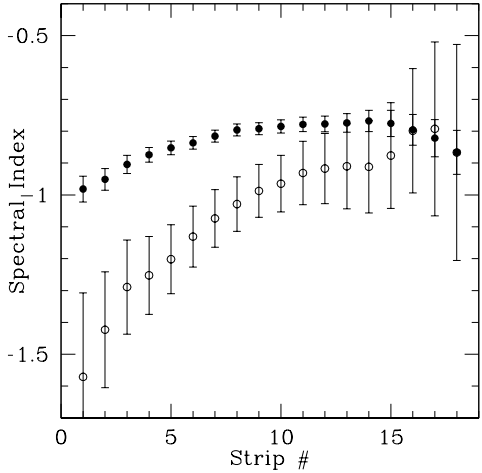}}
\caption{\footnotesize Spectral index trend across the radio relic
  region where strips run parallel to the long axis of the relics from
  the south-east edge (Strip 1) to the north-west edge (Strip
  18). Open points and errors show the spectral index between 1369
  and 1703 MHz, while the filled points and error bars show the 325 to
  1369 MHz spectral index.}
\label{A2256_slice}
\end{figure}

\cite{clarke06} found a spectral index trend between 1369 and 1703 MHz
across the Mpc scale relics in Abell 2256 such that the spectrum
steepens from the north-west toward the south-east. Comparison of this
to a similar spectral index map constructed between 325 MHz and 1369
MHz (Figure~\ref{A2256_slice}) reveals that the low and high-frequency
spectral indices are consistent near the north-west edge of the
relics. Interestingly though, the spectral index trend differs as one
moves to the south-east edge of the relics such that the low-frequency
spectral index is flatter by $\Delta \alpha \sim$ 0.5 on the
south-east edge of the relic compared to the high-frequency spectral
index in the same region. A plausible explanation for this spectral
index trend is that we are seeing the matching of the spectral index
values around $\alpha=-0.85$ near the current location of the outgoing
merger shock wave. Moving toward the south-east, in the direction from
which the shock originated, we are seeing the effects of synchrotron
and inverse Compton losses steepening the high-frequency spectral
index measurements.

If we assume that the above scenario is correct, then we can use the
spectral index measurement at the north-west edge of the relic to
place constraints on the Mach number of the shock. Assuming standard
Rankine-Hugonoit jump conditions in the post-shock gas, as in
\citet{1998A&A...332..395E}, we can relate the shock Mach number $M$ to the spectral
index $\alpha$ as
$$ M \sim \sqrt{2(2+\alpha_i)\over{1+2\alpha_i-3\gamma}},$$ where
$\alpha_i=1-2\alpha$ is the cosmic ray spectral index, and
$\gamma$=5/3 is the adiabatic index of the gas. For the observed
spectral index of $\alpha=-0.85$ at the north-west edge of the relic
we predict a Mach number of $M$=2.6. We find no clear evidence for an
X-ray shock in our combined XMM-Newton data (Figure~\ref{A2256_tmap}),
although \citet{sun} note the presence of a hot region to the north of
the main cluster which appears coincident with the eastern
relic. Deeper $Chandra$ and/or XMM-Newton observations are required to
search for the X-ray signatures of the predicted shock.

\section{X-ray Properties of the ICM }

We have combined 9 separate XMM-Newton data sets for both an imaging
and spectral study of Abell 2256. The images reveal similar ICM
substructure as seen in Chandra observations of \cite{sun} where the
main cluster and subcluster dominate the emission. 

\begin{figure}
\resizebox{\hsize}{!}{\includegraphics[clip=true]{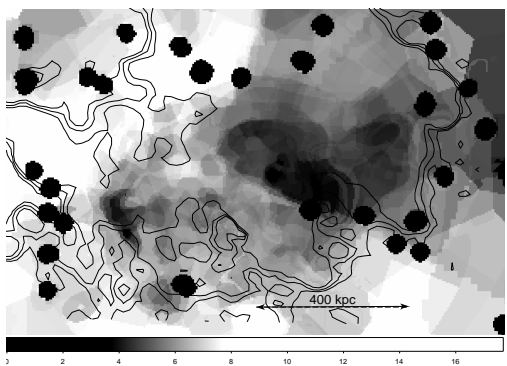}}
\caption{\footnotesize ICM temperature map of the central region of Abell 2256 determined
  from XMM-Newton data shown in greyscale with the other three radio
  contours \citet{clarke06} shown in black for
  reference. The cluster shows significant temperature structure,
  including two dark regions which trace the cool gas. }
\label{A2256_tmap}
\end{figure}

The spectral analysis was carried out using techniques of
\citet{2005A&A...442..827F}. Based on the Voroni tessellation maps
\citep{2006MNRAS.368..497D}, we selected roughly 100 contiguous
regions in each of 9 separate masks for spectral analysis. The regions
were chosen based on variations in either intensity or hardness
ratio. We averaged over the results from each of the 9 separate masks
to reduce the pixelization in the final images. Using these methods we
have constructed temperature, entropy ratio and pressure ratio
maps. We show in Figure~\ref{A2256_tmap} the resulting averaged
temperature map with the outer three 1.4 GHz radio contours of
\citet{clarke06} overlaid for reference.

The temperature map of the central region of the cluster shows
significant structure with clear regions of cool emission (dark areas)
embedded within hotter gas (bright areas). Overall the cluster core is
cool ($kT\sim 6-7$ keV) and is surrounded by hotter gas ($kT\sim 7-8$
keV), consistent with previous X-ray studies. The most striking
features of the temperature map are two regions of very cool gas
located to the north-west and south-east of the cluster center. Both
regions have temperatures of $\sim$ 4 keV and are clearly distinct
from the surrounding regions. 

The north-west cool region is associated with the merging subcluster
and sits at the inner edge of the radio relic region. The cool region
to the south-east ressembles a cold front and is co-incident with the
X-ray shoulder discussed by \citet{sun}. There appears to be a sharp
edge on the south-east portion of this region and a trail of cool gas
leading back toward the cluster core. The sharp edge and tail both
suggest the direction of motion of the south-east clump is toward the
south-east, similar in direction to the motion of the larger cold
front to the north-west. 

The entropy ratio maps shows two distinct minima in the cluster
associated with the merging subcluster and the shoulder. The complex
morphology seen for the subcluster in the temperature map is also
reflected in the entropy ratio map. The pressure ratio map, made using
the base pressure profiles of \citet{2005A&A...442..827F}, shows
evidence of a low pressure core co-incident with the radio halo. The
pressure ratio map shows a number of fluctuation at the $10-20\%$
level which are possibly associated with turbulence in the cluster.

\section{Summary}

Abell 2256 continues to play a central role in detailed studies of
merging clusters and the connection to diffuse radio
emission. Upcoming low-frequency observations with the EVLA, GMRT and
LOFAR will undoubtedly reveal additional details on this extraordinary
system. To fully explore this system it will also be important to
obtain further deep, high-resolution $Chandra$ and/or XMM-Newton
images of the cluster to undertake additional spectral analysis of
many of the complex features in this system.

\begin{acknowledgements}
Basic research in radio astronomy at the Naval Research Laboratory is
funded by 6.1 Base funding.
\end{acknowledgements}

\bibliographystyle{aa}

\end{document}